\documentclass[aps,showpacs,showkeys,superscriptaddress]{revtex4}

\usepackage{graphicx}

\begin{document}
\title{Electronic Structure of Disclinated Graphene in an
Uniform Magnetic Field}

\author{J. Smotlacha}\email{smota@centrum.cz}
\affiliation{Faculty of Nuclear Sciences and Physical Engineering, Czech Technical University, Brehova 7, 110 00 Prague,
Czech Republic}

\author{R. Pincak}\email{pincak@saske.sk}
\affiliation{Bogoliubov Laboratory of Theoretical Physics, Joint
Institute for Nuclear Research, 141980 Dubna, Moscow region, Russia}
\affiliation{Institute of Experimental Physics, Slovak Academy of Sciences,
Watsonova 47,043 53 Kosice, Slovak Republic}

\author{M. Pudlak}\email{pudlak@saske.sk}
\affiliation{Institute of Experimental Physics, Slovak Academy of
Sciences, Watsonova 47,043 53 Kosice, Slovak Republic}

\date{\today}

\pacs{73.22.Pr; 81.05.ue}

\keywords{graphene, heptagonal defect, elasticity, carbon nanohorns,
disclination}

\def\wu{\widetilde{u}}
\def\wv{\widetilde{v}}

\begin{abstract}
The electronic structure in the vicinity of the 1-heptagonal and
1-pentagonal defects in the carbon graphene plane is investigated.
Using a continuum gauge field-theory model the local density of
states around the Fermi energy is calculated for both cases. In
this model, the disclination is represented by an SO(2) gauge
vortex and corresponding metric follows from the elasticity properties of the graphene membrane. To enhance the interval of energies, a self-consistent
perturbation scheme is used. The Landau states are investigated and compared with the predicted values.
\end{abstract}

\maketitle

\section{Introduction}\

Nanostructured carbon materials are the materials with a special
geometrical structure of their molecules which we call carbon
nanoparticles. This geometrical structure is accompanied by
topological defects in a hexagonal plane lattice called graphene.
In most cases, these defects originate from the presence of the
pentagons for the positive curvature and the heptagons for the
negative curvature \cite{1}.\\

There are known variously shaped carbon nanostructures. The most
famous is the fullerene which has the structure of the soccer ball
and can be approximated as a sphere. It is composed of 60 carbon
atoms which create 20 hexagons and 12 pentagons \cite{2,kroto2}. However,
other structures exist, for example, nanocones, nanotoroids, nanotubes,
nanohorns etc. Wide varieties of electronic properties of these structures was studied. Some examples are given by magnetic properties \cite{o2, a28}, optical absorption spectra or electronic properties of nanotube caps \cite{y32}. These properties give a potential use in nanoscale devices like quantum wires, nonlinear electronic elements, transistors, molecular memory devices or electron field emitters. From the theoretical point of view, it was predicted and experimentally verified by scanning tunelling microscopy that metallic or semiconducting properties of carbon nanotubes depend on whether or not the difference $n-m$ of the components of the chiral vector $(n,m)$ is multiple of $3$ \cite{3,ando,crespi}.\\

 More complicated structures can arise when two variously shaped parts
of nanoparticles are connected by a region with pentagon-heptagon
pairs. It seems that the best approximation for
pentagonal and heptagonal areas is hyperboloid - positively curved
for pentagons and negatively curved for heptagons \cite{4}.\\

The electronic properties of these structures can be explored by
solving Dirac equation at a curved surface \cite{6}. In this
paper, after introducing the computational formalism, some
geometrical properties of the defects are researched. After doing
this, we research the Gaussian curvature of the surface. Then with
the help of the Dirac equation the local density of states (LDoS)
for disclinated areas near the Fermi level (close to the zero
energy) is calculated for both pentagonal and heptagonal defects.
It will be influenced by a uniform magnetic field. Then we compare the electronic properties of both models and, finally, we research the corresponding Landau states and compare them with the approximation formulas from the earlier works. The model described in \cite{4} is used.\\

\section{Basic formalism}\

First we introduce the Dirac equation in (2+1) dimensions. It has
the form
\begin{equation}\label{1}i\gamma^{\alpha}e_{\alpha}^{\mu}[\nabla_{\mu}-ia_{\mu}-iA_{\mu}]\psi=
E\psi.\end{equation} The wave function $\psi$, so-called bispinor,
is composed of two parts,
\begin{equation}\psi=\left(\begin{array}{c}\psi_A \\ \psi_B\end{array}\right),\end{equation}
each corresponding to different sublattices of the curved graphene
sheet. The gauge field $a_{\mu}, \mu=\xi,\varphi$ arises from spin
rotation invariance for atoms of different sublattices $A$ and $B$
in the Brillouin zone \cite{5}.

The zweibein $e_{\alpha}$ stands for incorporating fermions on the
curved 2D surface and it has to yield the same values of observed
quantities for different choices related by the local SO(2)
rotations,
\begin{equation}e_{\alpha}\rightarrow e_{\alpha}'=\Lambda^{\beta}_{\alpha}e_{\beta},\hspace{1cm}
\Lambda^{\beta}_{\alpha}\in SO(2).\end{equation} For this purpose,
 a covariantly constant local gauge field $\omega_{\mu}$ is incorporated \cite{diferm}:
\begin{equation}\partial_{\mu}e^{\alpha}_{\nu}-\Gamma^{\rho}_{\mu\nu}e^{\alpha}_{\rho}+
(\omega_{\mu})^{\alpha}_{\beta} e^{\beta}_{\nu}=0,\end{equation}
where $\Gamma_{\mu}$ is the Levi-Civita connection coming from the
metric $g_{\mu\nu}$ (see below). Then $\omega_{\mu}$ is called the
spin connection. Next, the covariant derivative $\nabla_{\mu}$ is
defined as
\begin{equation}\nabla_{\mu}=\partial_{\mu}+\Omega_{\mu},\end{equation}
where
\begin{equation}\Omega_{\mu}=\frac{1}{8}\omega^{\alpha\beta}_{\mu}[\gamma_{\alpha},\gamma_{\beta}]
\end{equation}
denotes the spin connection in the spinor representation. The
Dirac matrices $\gamma_{\alpha}$ can be replaced in two dimensions
by the Pauli matrices $\sigma_{\alpha}$:
\begin{equation}\gamma_1=-\sigma_2,\hspace{1cm}\gamma_2=\sigma_1.\end{equation}
$A_{\mu}$ is the vector potential arising from the external
magnetic field.\\

The metric $g_{\mu\nu}$ of the 2D surface comes from the
parametrization with the help of two parameters $\xi$, $\varphi$:
\begin{equation}\label{8}(\xi,\varphi)\rightarrow \overrightarrow{R}=(x(\xi,\varphi),y(\xi,\varphi),
z(\xi,\varphi)),\end{equation} where
\begin{equation}0\leq\xi<\infty,\hspace{1cm}0\leq\varphi<2\pi.\end{equation}
The 4 components of the metric are defined as
\begin{equation}g_{\mu\nu}=\partial_{\mu}\overrightarrow{R}\partial_{\nu}\overrightarrow{R}.
\end{equation}
The above-mentioned hyperboloid geometry has for both cases,
heptagons and pentagons, very similar but not the same
parametrization. We research it in separate chapters. The
non-diagonal components of the metric are
\begin{equation}g_{\xi\varphi}=g_{\varphi\xi}=0.\end{equation}

For the zweibeins and the diagonal components of metric holds
\begin{equation}e^1_{\xi}=\sqrt{g_{\xi\xi}}\cos{\varphi},\hspace{1cm}
e^1_{\varphi}=-\sqrt{g_{\varphi\varphi}}\sin{\varphi},\end{equation}
\begin{equation}e^2_{\xi}=\sqrt{g_{\xi\xi}}\sin{\varphi},\hspace{1cm}
e^2_{\varphi}=\sqrt{g_{\varphi\varphi}}\cos{\varphi},\end{equation}
and for the spin connection coefficients $\omega_{\mu}$
\begin{equation}de^1=-\omega^{12}\wedge e^2,\hspace{1cm}de^2=-\omega^{21}\wedge
e^1,\hspace{1cm}\omega^{12}=-\omega^{21},\end{equation} so
\begin{equation}\label{om1}\omega^{12}_{\varphi}=-\omega^{21}_{\varphi}=1-\frac{\partial_{\xi}\sqrt{g_{\varphi\varphi}}}
{\sqrt{g_{\xi\xi}}}=2\omega,
\end{equation}
\begin{equation}\label{om2}\omega^{12}_{\xi}=\omega^{21}_{\xi}=0.\end{equation}
Then the coefficients $\Omega_{\mu}$ are
\begin{equation}\Omega_{\xi}=0,\hspace{1cm}\Omega_{\varphi}=i\omega\sigma_3.\end{equation}
Substituting
\begin{equation}\left(\begin{array}{c}\psi_A \\ \psi_B\end{array}\right)
=\frac{1}{\sqrt[4]{g_{\varphi\varphi}}}\left(\begin{array}{c}\wu(\xi)e^{i\varphi
j}\\ \wv(\xi)e^{i\varphi(j+1)}\\\end{array}\right),\hspace{1cm} j=0,\pm
1,...\end{equation} and making up (\ref{1}) we get
\begin{equation}\label{r1}\partial_{\xi}\wu-(j+1/2-a_{\varphi}+A_{\varphi})
\sqrt{\frac{g_{\xi\xi}}{g_{\varphi\varphi}}}
\wu=E\sqrt{g_{\xi\xi}}\wv,\end{equation}
\begin{equation}\label{r2}-\partial_{\xi}\wv-(j+1/2-a_{\varphi}+A_{\varphi})
\sqrt{\frac{g_{\xi\xi}}{g_{\varphi\varphi}}}
\wv=E\sqrt{g_{\xi\xi}}\wu.\end{equation}

\section{Geometrical properties}\

To find the solution of (\ref{r1}), (\ref{r2}), the knowledge of the
influence of the defects on the components of the
 metric $g_{\mu\nu}$ is needed. It is characterized by the Frank index
  $\nu$ which depends on the number of the defects.\\

It is possible to try to approximate the geometry by the metric of the cone \cite{crespi}. But this procedure causes discrepancy in the calculation of the corresponding gauge flux \cite{4}. Here is suggested a method, how to avoid this problem.\\

\subsection{Heptagonal defects}\

In case of the negative curvature and affiliated heptagonal
defects, in (\ref{8}) the parametrization seems to be
\begin{equation}\label{paramP0}(\xi,\varphi)
\rightarrow(a\cosh\xi\cos\varphi,a\cosh\xi\sin\varphi,c\sinh\xi),\end{equation}
where $a$ and $c$ are some dimensionless parameters. The
corresponding diagonal components of the metric are
\begin{equation}\label{metr}g_{\xi\xi}=a^2\sinh^2\xi+c^2\cosh^2\xi,\hspace{1cm}
g_{\varphi\varphi}=a^2\cosh^2\xi\end{equation} and the nonzero spin
connection term
\begin{equation}\label{spin}\omega^{12}_{\varphi}=
1-\frac{a\sinh\xi}{\sqrt{g_{\xi\xi}}}.\end{equation} The defect
arises by the so-called cut and glue procedure - we cut a line in
the graphene plane, add a $60^{\circ}$ area and glue the arising
borders \cite{crespi}. The geometrical properties of the new surface can be
described with the help of the gauge potentials
$\overrightarrow{W}_{\mu}^{(0)}$ which change the initial
components of the metric (now denoted $g_{\mu\nu}^{(0)}$)
\cite{8}:
\begin{equation}g_{\mu\nu}^{(0)}\rightarrow g_{\mu\nu}=\nabla_{\mu}\overrightarrow{R}_{(0)}\cdot
\nabla_{\nu}\overrightarrow{R}_{(0)},\end{equation} where
\begin{equation}\nabla_{\mu}\overrightarrow{R}_{(0)}=\partial_{\mu}\overrightarrow{R}_{(0)}+
[\overrightarrow{W}_{\mu}^{(0)},\overrightarrow{R}_{(0)}].\end{equation}
Then
\[g_{\mu\nu}=\partial_{\mu}\overrightarrow{R}_{(0)}\cdot
\partial_{\nu}\overrightarrow{R}_{(0)}+\partial_{\mu}\overrightarrow{R}_{(0)}
[\overrightarrow{W}_{\nu}^{(0)},\overrightarrow{R}_{(0)}]+\]
\begin{equation}+\partial_{\nu}\overrightarrow{R}_{(0)}[\overrightarrow{W}_{\mu}^{(0)},
\overrightarrow{R}_{(0)}]+(\overrightarrow{W}_{\mu}^{(0)}\overrightarrow{W}_{\nu}^{(0)})
\overrightarrow{R}_{(0)}^2-(\overrightarrow{W}_{\mu}^{(0)}\overrightarrow{R}_{(0)})
(\overrightarrow{W}_{\nu}^{(0)}\overrightarrow{R}_{(0)})\end{equation}
and the components of the metric and the spin connection term will
be changed in the way that
\begin{equation}g_{\xi\xi}=a^2\sinh^2\xi+c^2\cosh^2\xi,\hspace{1cm}
g_{\varphi\varphi}=a^2\alpha^2\cosh^2\xi,\end{equation}
\begin{equation}\label{spin1}\omega^{12}_{\varphi}=
1-\frac{a\alpha\sinh\xi}{\sqrt{g_{\xi\xi}}},\hspace{1cm}\alpha=1+\nu,\end{equation}
where $\nu=N/6$ is called the Frank index and $N$ is the number of
heptagons in the defect. In this paper, we take $N=1$. Let us stress that for the higher number of defects, the geometrical structure is more complicated and we have to take into account next assumptions \cite{crespi,berber}.\\

 We can
encircle the origin of the defect ($\xi=0$) by a closed loop
$C_{\epsilon}$ and integrate over it. The result is
\begin{equation}\oint_{C_{\epsilon}}d\overrightarrow{s}=2\pi\nu.\end{equation}\\
 No transformation of variables can change this value. If the
values of the gauge field $\overrightarrow{W}_{\mu}^{(0)}$ are
\begin{equation}\label{w1}W_{\mu}^{(0)i=1,2}=0,\hspace{1cm}W_{\mu}^{(0)i=3}=W_{\mu}^{(0)},\end{equation}
where
\begin{equation}\label{w2}W_x^{(0)}=-\nu y/r^2,\hspace{1cm}W_y^{(0)}=\nu x/r^2,\hspace{1cm}
r=\sqrt{x^2+y^2},\end{equation} then
\begin{equation}\oint_{C_{\epsilon}}d\overrightarrow{s}=2\pi\nu=\oint_{C_{\epsilon}}
W_{\mu}^{(0)}dx^{\mu},\end{equation} so
$\overrightarrow{W}_{\mu}^{(0)}$ serves as a vortex-like potential
with a nonzero flux. This flux should be eliminated by the
corresponding integral over the spin connection, so we must get
\begin{equation}\label{fl}\lim\limits_{\epsilon\rightarrow 0}
\oint_{C_{\epsilon}}\omega^{12}_{\varphi}
d\varphi=-2\pi\nu.\end{equation} Substituting (\ref{spin1}) into
the appropriate integral, the required result is really
obtained.\\

For our purpose, the gauge field $a_{\varphi}=N/4$. In the general case, $a_{\varphi}$ depends on two constants, $N$ and $M$ as $a_{\varphi}=N/4+M/3$, where $M=-1,0,1$ for an even number of defects and $M=0$ for an odd number of defects \cite{crespi,4,5}.\\

If the magnetic field is chosen
in such a way that $\overrightarrow{A}=B(y,-x,0)/2$, then
\begin{equation}\label{HeptA} A_{\varphi}=-\Phi\cosh^2\xi,\hspace{1cm}A_{\xi}=0,\end{equation}
where
\begin{equation}\label{Phi}\Phi=\frac{1}{2}a^2\Phi_0 B,\hspace{1cm}\Phi_0=\frac{e}{\hbar c}.
\end{equation}
Geometric units are used, i.e. $e=\hbar=c=1.$\\

\subsection{Pentagonal defects}\

The case of the positive curvature is described in more detail in \cite{4}. The parametrization
 is
changed into
\begin{equation}\label{paramH}(\xi,\varphi)
\rightarrow(a\sinh\xi\cos\varphi,a\sinh\xi\sin\varphi,c\cosh\xi),\end{equation}
and the diagonal components of the metric
\begin{equation}g_{\xi\xi}=a^2\cosh^2\xi+c^2\sinh^2\xi,\hspace{1cm}
g_{\varphi\varphi}=a^2\sinh^2\xi.\end{equation}\\

 Introducing the
gauge potentials $\overrightarrow{W}_{\mu}^{(0)}$ as for the
heptagonal defects, the component $g_{\varphi\varphi}$ of the
metric changes in the way that
\begin{equation}g_{\varphi\varphi}=a^2\alpha^2\sinh^2\xi,\end{equation}
where $\alpha=1-\nu$. This means that in the cut and glue
procedure we cut a $60^{\circ}$ area instead of insertion. Then the
nonzero spin connection term is
\begin{equation}\omega^{12}_{\varphi}=
1-\frac{a\alpha\cosh\xi}{\sqrt{g_{\xi\xi}}}.\end{equation}

The values of the gauge field and the magnetic field are the same as
in the previous case,
\begin{equation}a_{\varphi}=N/4,\hspace{1cm}\overrightarrow{A}=B(y,-x,0)/2,\end{equation}
so for the parametrization chosen
\begin{equation}\label{PentA} A_{\varphi}=-\Phi\sinh^2\xi,\hspace{1cm}A_{\xi}=0,\end{equation}
where $\Phi$ is defined as in (\ref{Phi}).\\

\section{Curvature}\

The Gaussian curvature is defined as

\begin{equation}\label{curv}K=\frac{(\partial_{xx}f)(\partial_{yy}f)-(\partial_{xy}f)^2}
{(1+(\partial_xf)^2+(\partial_yf)^2)^2},\end{equation}
where $f$ means the $z$ coordinate in (\ref{8}) expressed with the help of $x$ and $y$, i.e. formally we take $f(x,y)=z(\xi,\varphi)$. According to our presumptions, this quantity should be negative for heptagonal defects and positive for pentagonal defects.\\

\subsection{Heptagonal defects}\

Comparison of (\ref{8}) and (\ref{paramP0}) implicates
\begin{equation}\frac{x^2+y^2}{a^2}-\frac{f(x,y)^2}{c^2}=1,\end{equation}
so
\begin{equation}f(x,y)=\frac{c}{a}\sqrt{x^2+y^2-a^2}=\frac{c}{a}\sqrt{r^2-a^2}.\end{equation}
After making up derivations $\partial_{xx}f$, $\partial_{yy}f$ and $\partial_{xy}f$ we get
\begin{equation}K=-\frac{c^2}{(r^2(1+\eta)-a^2)^2},\end{equation}
where we use the initial definition of $\eta$, given in (\ref{32}). As we clearly see, this expression is negative for arbitrary values of $r$.\\

\subsection{Pentagonal defects}\

For the parametrization (\ref{paramH}), we get
\begin{equation}\frac{f(x,y)^2}{c^2}-\frac{x^2+y^2}{a^2}=1,\end{equation}
so
\begin{equation}f(x,y)=\frac{c}{a}\sqrt{x^2+y^2+a^2}=\frac{c}{a}\sqrt{r^2+a^2}.\end{equation}
Making up the corresponding derivations, we get using (\ref{curv})
\begin{equation}K=\frac{c^2}{(r^2(1+\eta)+a^2)^2},\end{equation}
so the Gaussian curvature is all the time positive for the pentagonal defects.\\

\section{Solution of the Dirac equation}\

The solution of (\ref{r1}),(\ref{r2}) for heptagonal and
pentagonal defects is introduced and the local density of states
is calculated here. The linear elasticity theory \cite{8, EP28} is used. For the numerical calculations of LDoS, method described in \cite{7} is exploited.

\subsection{Heptagonal defects}\

 The form of (\ref{r1}),(\ref{r2}) will be
\begin{equation}\label{eq1H}\partial_{\xi}\wu-(\widetilde{j}-\widetilde{\Phi}\cosh^2\xi)
\sqrt{\tanh^2\xi+\eta}\wu=E\sqrt{g_{\xi\xi}}\wv,\end{equation}
\begin{equation}\label{eq2H}-\partial_{\xi}\wv-(\widetilde{j}-\widetilde{\Phi}\cosh^2\xi)
\sqrt{\tanh^2\xi+\eta}\wv=E\sqrt{g_{\xi\xi}}\wu,\end{equation}
where
\begin{equation}\label{32}\widetilde{j}=(j+1/2-a_{\varphi})/\alpha,\hspace{1cm}
\widetilde{\Phi}=\Phi/\alpha,\hspace{1cm}\eta=c^2/a^2.\end{equation}
$\eta\ll1$ is a dimensionless parameter which contains the elasticity properties of the initial graphene plane. Due to these properties, the defects can be interpreted as small perturbations in that graphene plane. In case of finite elasticity, we can do an approximation $\eta\sim\sqrt{\nu\epsilon}$, where $\epsilon\leq 0.1$
\cite{4}. By this way, the elasticity is described by a small parameter $\epsilon$. Its value is usually taken between $0.01$ and $0.1$. If we do some neccesary corrections of the gauge field $\omega_{\mu}$, for $\epsilon\rightarrow 0$ we get the metric of the cone.\\

Suppose now $E=0$. This energy corresponds to so-called zero-energy mode which is appropriate for the electron states at the Fermi level. Then, the solution of (\ref{eq1H}),(\ref{eq2H}) is
\begin{equation}\label{52}\wu_0(\xi)=C(\triangle(\xi)+k\sinh\xi)^{k\widetilde{j}-
\frac{\eta\widetilde{\Phi}}{2k}}\left(\frac{\cosh\xi}{\triangle(\xi)+\sinh\xi}\right)^{\widetilde{j}}
\exp\left(-\frac{\widetilde{\Phi}\triangle(\xi)\sinh\xi}{2}\right),\end{equation}
\begin{equation}\label{53}\wv_0(\xi)=C'(\triangle(\xi)+k\sinh\xi)^{-k\widetilde{j}+
\frac{\eta\widetilde{\Phi}}{2k}}\left(\frac{\cosh\xi}{\triangle(\xi)+\sinh\xi}\right)^{-\widetilde{j}}
\exp\left(\frac{\widetilde{\Phi}\triangle(\xi)\sinh\xi}{2}\right),\end{equation}
where
\begin{equation}k=\sqrt{1+\eta},\hspace{1cm}\triangle(\xi)=\sqrt{k^2\cosh^2\xi-1}\end{equation}
and $C$, $C'$ are the normalization constants.

For the nonzero values of $E$ the solution can be written as in
\cite{7}
\begin{equation}\label{nonf}\wu(\xi)=\wu_0(\xi)\mathcal{U}(\xi),\hspace{1cm}
\wv(\xi)=\wv_0(\xi)\mathcal{V}(\xi),\end{equation} where
\begin{equation}\mathcal{U}(\xi)=\mathcal{U}^{(0)}(\xi)+\varepsilon\mathcal{U}^{(1)}(\xi)+\cdots+
\varepsilon^n\mathcal{U}^{(n)}(\xi)\end{equation} and
\begin{equation}\mathcal{V}(\xi)=\mathcal{V}^{(0)}(\xi)+\varepsilon\mathcal{V}^{(1)}(\xi)+\cdots+
\varepsilon^n\mathcal{V}^{(n)}(\xi),\end{equation}
$\varepsilon=\frac{Ea}{\hbar v_F}$ and we take $\hbar=v_F=1$. Here $n$ is an integer number and it is chosen according to our requirements on the precision.
After substitution of this approximation into (\ref{eq1H}) and (\ref{eq2H}) we get
\begin{equation}\partial_{\xi}\mathcal{U}=\varepsilon\triangle(\xi)\mathcal{V}\frac{\wv_0}{\wu_0},
\hspace{1cm}\partial_{\xi}\mathcal{V}=-\varepsilon\triangle(\xi)\mathcal{U}\frac{\wu_0}{\wv_0}.\end{equation}
Putting $\mathcal{U}^{(0)}=1$, $\mathcal{V}^{(0)}=0$ and $i=0, 1,..., n-1,$ in a numerical way, the form of the solution
will be
\begin{equation}\mathcal{U}^{(i+1)}(\xi)=\int\limits_0^{\xi}\mathcal{V}^{(i)}(\zeta)\Delta(\zeta)
\frac{\wv_0(\zeta)}{\wu_0(\zeta)}d\zeta,\end{equation}
\begin{equation}\label{nonl}\mathcal{V}^{(i+1)}(\xi)=-\int\limits_0^{\xi}\mathcal{U}^{(i)}(\zeta)
\Delta(\zeta)
\frac{\wu_0(\zeta)}{\wv_0(\zeta)}d\zeta.\end{equation}

The local density of states is for given $\xi_0$ defined as
\begin{equation}LDoS(E)=\wu^2(E,\xi_0)+\wv^2(E,\xi_0).\end{equation} For its
evaluation we have to calculate the normalization constants
$C$,$C'$. They differ for different values of $E$. For unnormalized
solutions $\wu'(\xi), \wv'(\xi)$ of (\ref{eq1H}),(\ref{eq2H}) and
each value of energy,
\begin{equation}\label{norm}
1/C^2=1/C'^2=\int\limits_0^{\xi_{max}}(\wu'(\xi)^2+\wv'(\xi)^2)d\xi.\end{equation}
Since for $\xi_{max}=\infty$, the integral diverges, some finite
value of $\xi_{max}$ in some interval, which is of particular
interest, is needed. In this work, we take $\xi_{max}=2.5$ and $\xi_{max}=2$. It follows from the parametrization (\ref{paramP0}) that for given $\xi$, the corresponding distance is $r=a\cosh\xi$, which means that for $a=1A$ we have $r_{max}=6.13A$, resp. $r_{max}=3.76A$. These values are of the same order as the size of the Brillouin zone which is formed by the single hexagons. Each atom in the hexagon lies at the distance $1.42A$ from its nearest neighbours \cite{wallace,slon}. This is the main principle of the tight-binding approximation \cite{mvoz} in which we only admit the influence of the nearest neighbours.

\subsection{Pentagonal defects}\

The form of (\ref{r1}),(\ref{r2}) is
\begin{equation}\label{eq1}\partial_{\xi}\wu-(\widetilde{j}-\widetilde{\Phi}\sinh^2\xi)
\sqrt{\coth^2\xi+\eta}\wu=E\sqrt{g_{\xi\xi}}\wv,\end{equation}
\begin{equation}\label{eq2}-\partial_{\xi}\wv-(\widetilde{j}-\widetilde{\Phi}\sinh^2\xi)
\sqrt{\coth^2\xi+\eta}\wv=E\sqrt{g_{\xi\xi}}\wu.\end{equation}

In case of $E=0$, the corresponding solution is
\begin{equation}\label{65}\wu_0(\xi)=C(\triangle(\xi)+k\cosh\xi)^{k\widetilde{j}+
\frac{\eta\widetilde{\Phi}}{2k}}\left(\frac{\sinh\xi}{\triangle(\xi)+\cosh\xi}\right)^{\widetilde{j}}
\exp\left(-\frac{\widetilde{\Phi}\triangle(\xi)\cosh\xi}{2}\right),\end{equation}
\begin{equation}\label{66}\wv_0(\xi)=C'(\triangle(\xi)+k\cosh\xi)^{-k\widetilde{j}-
\frac{\eta\widetilde{\Phi}}{2k}}\left(\frac{\sinh\xi}{\triangle(\xi)+\cosh\xi}\right)^{-\widetilde{j}}
\exp\left(\frac{\widetilde{\Phi}\triangle(\xi)\cosh\xi}{2}\right),\end{equation}
where
\begin{equation}k=\sqrt{1+\eta},\hspace{1cm}\triangle(\xi)=\sqrt{k^2\sinh^2\xi+1}.\end{equation}

For calculating the solution for nonzero values of $E$ and the
local density of states we use the same procedure as for the
heptagonal defects.\\

\subsection{Local density of states}\

\begin{figure}
{\includegraphics{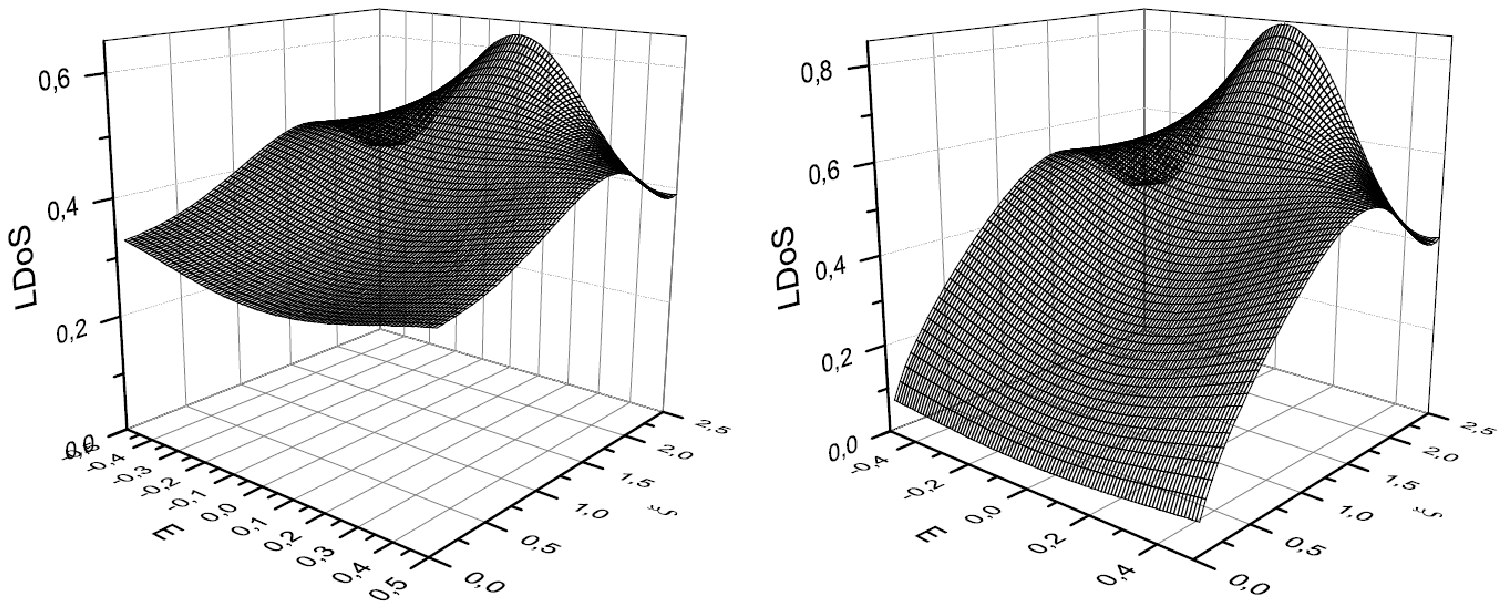}} \setcounter{figure}{0} \caption{LDoS
as a function of $E\in (-0.5,0.5)$ and $\xi\in (0,2.5)$ for
1-heptagon defects (left part) and 1-pentagon defects (right part)
for $B=0$;  $\epsilon=0.01$.}
\end{figure}

\begin{figure}
{\includegraphics{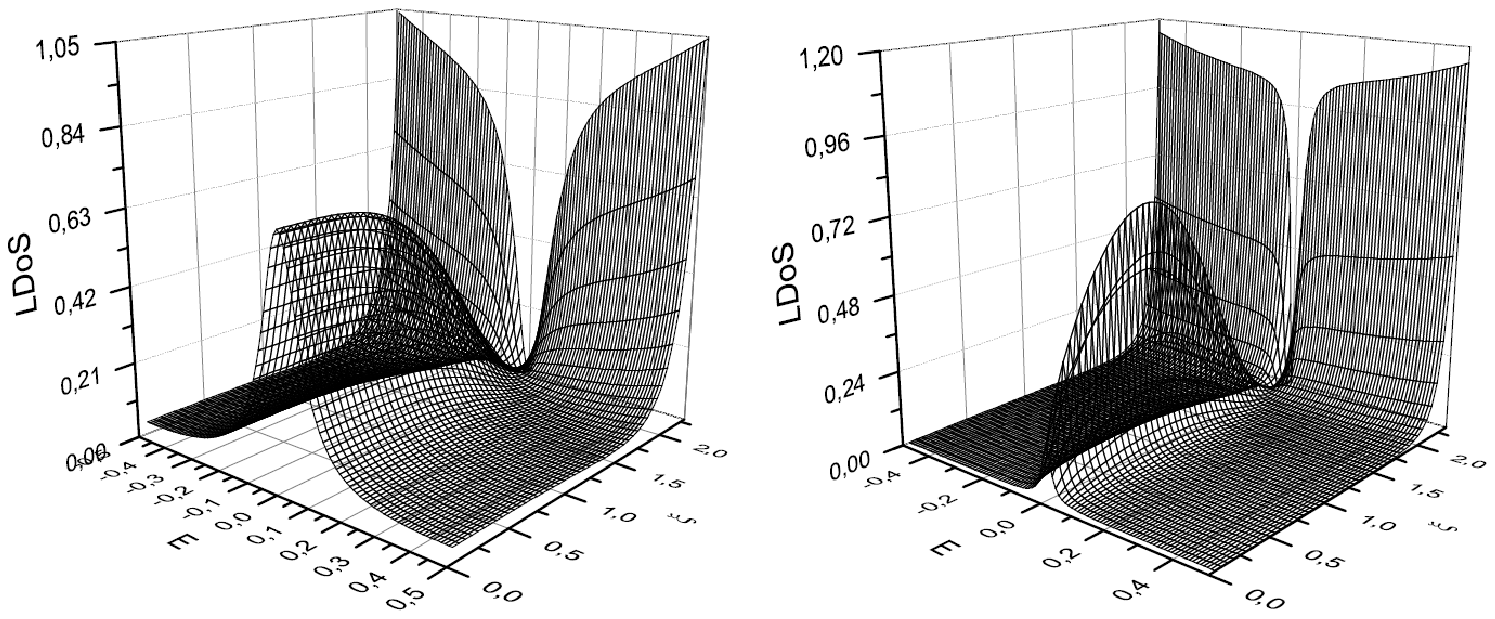}} \setcounter{figure}{1} \caption{LDoS
as a function of $E\in (-0.5,0.5)$ and $\xi\in (0,2.5)$ for
1-heptagon defects (left part) and 1-pentagon defects (right part)
for $B=0.5$;  $\epsilon=0.01$.}
\end{figure}

\begin{figure}
{\includegraphics{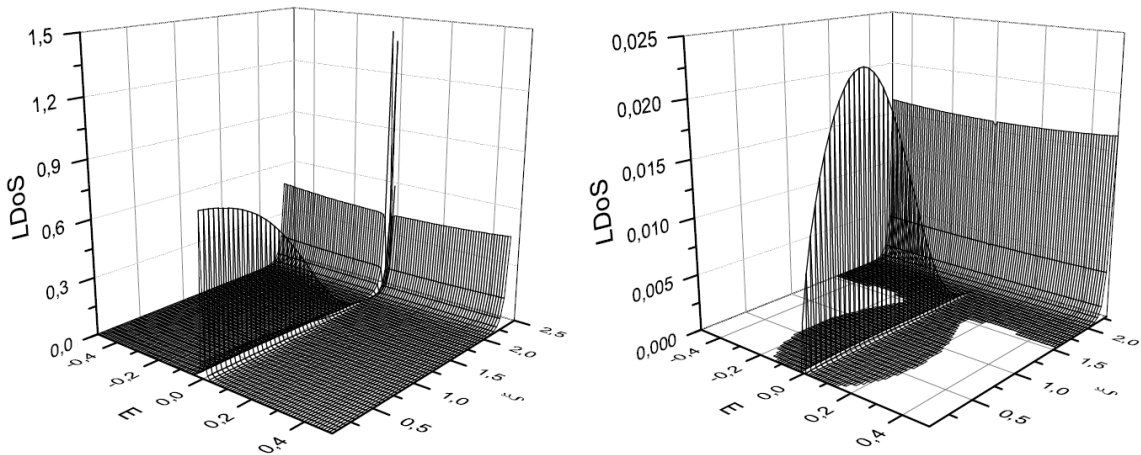}} \setcounter{figure}{2} \caption{LDoS
as a function of $E\in (-0.5,0.5)$ and $\xi\in (0,2.5)$ for
1-heptagon defects (left part) and 1-pentagon defects (right part)
for $B=1$;  $\epsilon=0.01$.}
\end{figure}

In Figs. 1-3, LDoS as a function of energy and the parameter
$\xi$ is presented for the surface with the defects formed by $1$ polygon. In all these figures, we choose $j=0$ in
(\ref{32}) and $\epsilon=0.01$ in the expression for $\eta$. We see the evidence that for growing $B$ or $\xi_{max}$, the LDoS is decreasing and the decrease is faster for the pentagonal defects. If we took larger $\xi_{max}$, LDoS would go to zero with an exception of a small number of energies for which we would get plane waves. But larger values of $\xi$ are unphysical because of limited interval of validity of the tight-binding approximation.

The values we chose enable us to compare LDoS for both kinds of defects for a small perturbation, where the difference between both approximations is not too large. Here we chose $\epsilon=0.01$, but there aren't any significant changes for the LDoS if we let $\epsilon$ to grow up to $0.1$, as we can easily see from the graphs in Figs. 4 and 5, where we compare LDoS for different values of magnetic field at a particular distance for $\epsilon=0.01$ and $0.1$.\\

\begin{figure}
{\includegraphics{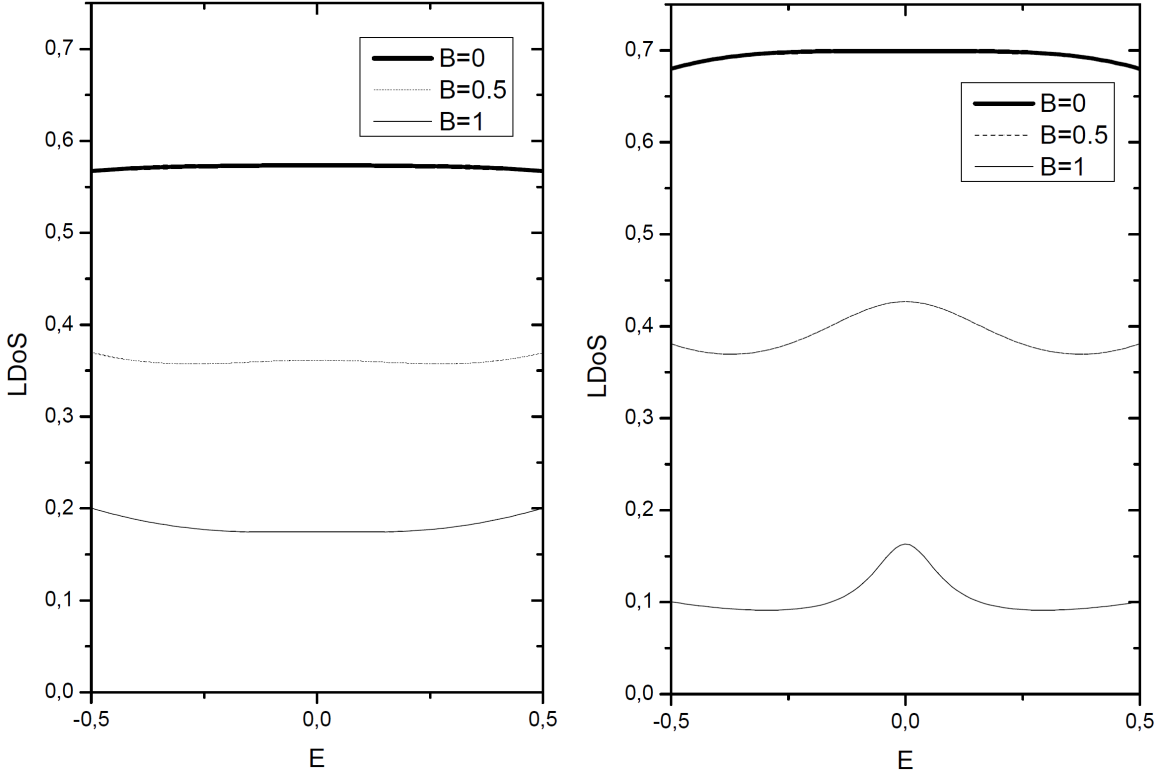}} \setcounter{figure}{3} \caption{LDoS as
a function of $E\in (-0.5,0.5)$ for 1-heptagon defects (left part)
and 1-pentagon defects (right part); various values of $B$ are used,
$\xi=1.5$, $\xi_{max}=2$;  $\epsilon=0.01$.}
\end{figure}

\begin{figure}
{\includegraphics{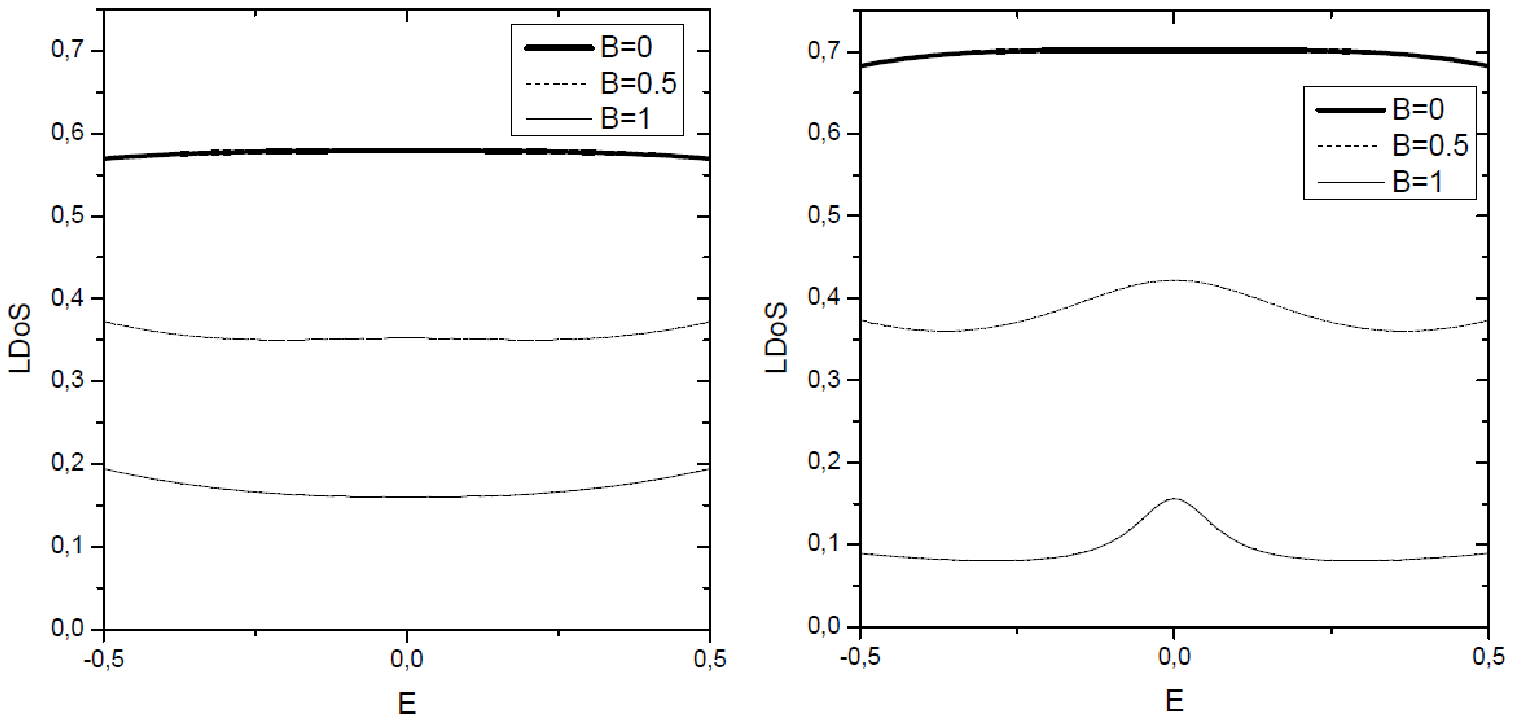}} \setcounter{figure}{4} \caption{LDoS
as a function of $E\in (-0.5,0.5)$ for 1-heptagon defects (left
part) and 1-pentagon defects (right part); various values of $B$ are
used, $\xi=1.5$, $\xi_{max}=2$; $\epsilon=0.1$.}
\end{figure}

\section{Landau states}\

In \cite{crespi} and \cite{4}, the Landau states for the researched defects are calculated for the conical and the hyperboloidal geometry for the positive curvature.  In case of the conical geometry, where the corresponding parametrization is
\begin{equation}(r,\varphi)
\rightarrow(r\cos\varphi,r\sin\varphi,c r)\end{equation}
($c$ is an arbitrary constant), we have, in agreement with the notation of \cite{crespi}, $2$ kinds of the Landau states.
Let us first denote
\begin{equation}\label{nu}\nu=\frac{6}{5}j\pm\frac{3}{10},\hspace{1cm} j=\pm\frac{1}{2},\pm\frac{3}{2},\pm\frac{5}{2},...\end{equation}

Then the first kind coincides with the Landau states of the planar graphene \cite{mcclure}. It corresponds to $\nu\geq0$ and the energy levels are
\begin{equation}E_n=\pm\sqrt{2B n},\hspace{1cm} n=0,1,2,...,\end{equation}
where $B$ means the magnetic field.

The second kind coincides with $\nu\leq0$, the energy levels are
\begin{equation}E_n=\pm\sqrt{2B}\sqrt{n-\nu+\frac{1}{2}},\hspace{1cm} n=0,1,2,...\end{equation}
and the list of these Landau states for different values of the magnetic field is presented in Tables $1$ and $2$. Because $j=-\frac{1}{2}$ in (\ref{nu}) corresponds \cite{crespi} to $j=0$ in (\ref{r1}), (\ref{r2}), we are looking for the corresponding energy levels.

\begin{table}

\caption{Landau states for pentagonal defects for $\nu\leq0$, $B=0.5$ and $j=-\frac{1}{2}$}
\begin{tabular*}{0.3\textwidth}{@{\extracolsep{1cm}}c c c}
\hline
& $n$ & $E_n$ \\
\hline
& $0$ & $\pm$ 0.89, $\pm$ 1.18 \\
& $1$ & $\pm$ 1.34, $\pm$ 1.55 \\
& $2$ & $\pm$ 1.67, $\pm$ 1.84 \\
& $3$ & $\pm$ 1.95, $\pm$ 2.10 \\
\hline
\end{tabular*}

\end{table}

\begin{table}

\caption{Landau states for pentagonal defects for $\nu\leq0$, $B=1$ and $j=-\frac{1}{2}$}
\begin{tabular*}{0.3\textwidth}{@{\extracolsep{1cm}}c c c}
\hline
& $n$ & $E_n$ \\
\hline
& $0$ & $\pm$ 1.25, $\pm$ 1.66 \\
& $1$ & $\pm$ 1.89, $\pm$ 2.19 \\
& $2$ & $\pm$ 2.35, $\pm$ 2.59 \\
& $3$ & $\pm$ 2.75, $\pm$ 2.96 \\
\hline
\end{tabular*}

\end{table}

Let us compare the Landau states for the hyperboloidal geometry with
the values calculated for cone and graphene. For this purpose, we do
an extension of the interval of energies for which we calculate the
LDoS. The result we see in Fig. 6. We see that for pentagonal
defects, the values $\pm 1.67$ and $\pm 2.19$ from Tables $1$ and
$2$ can be found by this method. The reason for the presence or
absence of other peaks could be the incompleteness of the list of
Landau states for hyperboloidal geometry and a low magnitude of some
peaks. It is also possible that some of Landau states characterizing
the conical and planar geometry don't appear in case of the
hyperboloidal geometry. For the negative curvature, the appropriate
energy levels have similar position but are shifted to the left.

Comparison of our results with the Landau states as expected for multilayer graphene \cite{koshino} is also interesting. These results differ in the way that for zero-energy states ($n=0$, see \cite{koshino}), there are nonzero Landau states with low magnitude. For higher $n$, the Landau states are calculated with the help of the approximation formula (using the notation of \cite{koshino})

\begin{equation}E^{(m)}_{n\geq 1,\pm}=\left(-\cos\frac{m\pi}{N+1}+\frac{2}{N+1}\cos^2\frac{m\pi}{2(N+1)}\right)\gamma_2\pm
\sqrt{\frac{1-(-1)^N}{N+1}\gamma_2^2\cos^4\frac{m\pi}{2(N+1)}+\frac{n(n+1)\Gamma_B^4}{4\gamma_1^2\sin^2\frac{m\pi}{2(N+1)}}},
\end{equation}
where $\Gamma_B=\sqrt{\frac{3}{2}B}\gamma_0$, $N$ is the number of the graphene layers and $\gamma_0=3$, $\gamma_1=0.4$, $\gamma_2=-0.02$. As we can see in Tables $3$ - $5$, the results acquired using this formula are higher in order then the Landau states calculated in this paper for the hyperboloid geometry. So, the Landau states for multilayer graphene and for the hyperboloid geometry are completely different. The reason may consist in considering the influence of atoms in neighbor and next-neighbor layers in the multilayer graphene which don't exist in case of the one-layer graphene.\\

\begin{table}

\caption{Landau states for $3$-layer planar graphene, $m=2$}
\begin{tabular*}{0.5\textwidth}{@{\extracolsep{1.5cm}}c c c c}
\hline
& $n$ & \multicolumn{2}{c}{$E_n$} \\
&     & $B=0.5$ & $B=1$ \\
\hline
& $0$ & $\pm$ 16.87 & $\pm$ 33.75 \\
& $1$ & $\pm$ 29.23 & $\pm$ 58.46 \\
& $2$ & $\pm$ 41.33 & $\pm$ 82.67 \\
\hline
\end{tabular*}

\end{table}

\begin{table}

\caption{Landau states for $4$-layer planar graphene, $m=3$}
\begin{tabular*}{0.5\textwidth}{@{\extracolsep{1.5cm}}c c c c}
\hline
& $n$ & \multicolumn{2}{c}{$E_n$} \\
&     & $B=0.5$ & $B=1$ \\
\hline
& $0$ & $\pm$ 14.75 & $\pm$ 29.50 \\
& $1$ & $\pm$ 25.55 & $\pm$ 51.10 \\
& $2$ & $\pm$ 36.13 & $\pm$ 72.26 \\
\hline
\end{tabular*}

\end{table}

\begin{table}

\caption{Landau states for $5$-layer planar graphene, $m=4$}
\begin{tabular*}{0.5\textwidth}{@{\extracolsep{1.5cm}}c c c c}
\hline
& $n$ & \multicolumn{2}{c}{$E_n$} \\
&     & $B=0.5$ & $B=1$ \\
\hline
& $0$ & $\pm$ 13.78 & $\pm$ 27.56 \\
& $1$ & $\pm$ 23.87 & $\pm$ 47.73 \\
& $2$ & $\pm$ 33.75 & $\pm$ 67.50 \\
\hline
\end{tabular*}

\end{table}

\begin{figure}
{\includegraphics{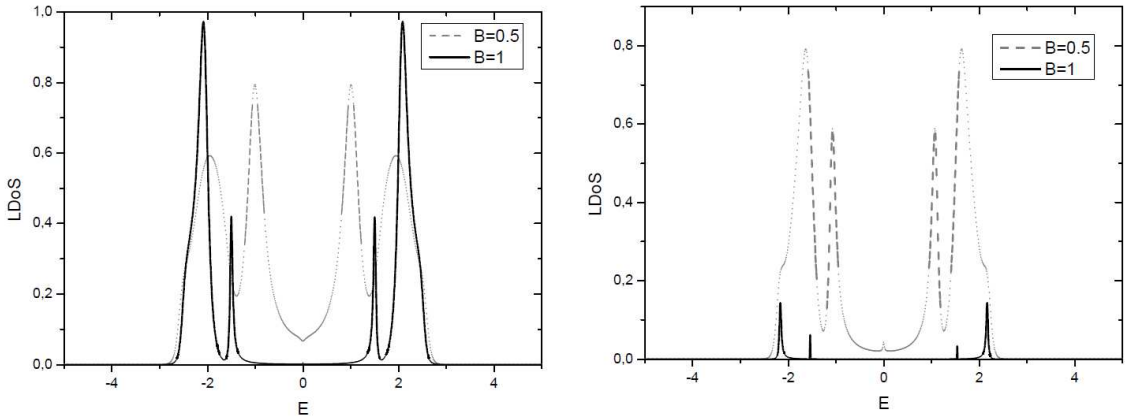}} \setcounter{figure}{5}
\caption{Landau states for heptagonal (left part) and pentagonal
defects (right part) for various values of $B$; $\xi_{max}=2$,
$\xi=2$, $\epsilon=0.01$.}\end{figure}

\section{Conclusion}\

We have studied the electronic structure of disclinated graphene in the vicinity of heptagonal and pentagonal defects
depending on the kind of the curvature (negative or positive).
Hyperboloidal parametrizations (\ref{paramP0}),(\ref{paramH})  were assumed after rejection of the conical metric. The continuum
field-theory gauge model was used in which the disclinations are
incorporated using the vortex-like potential (\ref{w1}),
(\ref{w2}) for calculation the components of the metric. The arising fictious flux was compensated by the gauge flux of spin connection field (\ref{om1}),(\ref{om2}). The potential (\ref{w1}),
(\ref{w2}) also results in the dependence of the corresponding Dirac equation on
the Frank index $\alpha$ which includes the number of defects. The defects are involved in (\ref{32}) with help of the parameter $\epsilon$ which comes from the elasticity properties of the graphene.

Next we incorporated an uniform magnetic field (\ref{HeptA}),(\ref{PentA}) which can significantly influence LDoS. It is calculated from the solution of the Dirac equation which we get numerically with help of the extension of an analytical solution for zero-energy modes (\ref{52}),(\ref{53}),(\ref{65}),(\ref{66}).

In all presented figures, the behaviour of the LDoS is compared for
both kinds of defects. For very small values of $\epsilon$, this
behaviour  is for both kinds of defects similar, but it is more
spread for heptagonal defects. As we found, the varying value of
$\epsilon$ does not affect LDoS significantly. It shows that with a
little change of elasticity the change of LDoS is also very small.
It could be done by big rigidity of graphene structure as well.
We also compared the resulting Landau states with the theoretical prediction coming from the corresponding values for graphene plane and conical metric. We see that the different geometrical structure influences the position of the Landau states.\\

To conclude, the presented results could have a large potential use
for next calculations of metallic properties of carbon
nanostructures which serve as a wide spectra of electronic devices.
Let us mention the significance of the zero-energy modes. Generally,
they always appear as a solution for disclinated graphene in
presence of the magnetic field \cite{jackiw} and they play a key
role in explanation of anomalies, paramagnetism, high-temperature
superconductance etc.

We have to stress that we assumed defects in which appeared only $1$ heptagon or $1$ pentagon. For the higher number of polygons in defects, the calculation is more complicated especially for heptagons, because in contrast to pentagonal defects, problems with the geometrical interpretation occur. It is useful to do next calculations for more complicated forms of defects.

\vskip 0.2cm ACKNOWLEDGEMENTS --- We very thanks Prof. V.A.Osipov
for the helpful comments and advice. The work was supported by the
Slovak Academy of Sciences in the framework of CEX NANOFLUID, and by
the Science and Technology Assistance Agency under Contract No. APVV
0509-07 , 0171 10,  VEGA Grant No. 2/0069/10 and Ministry of
Education Agency for Structural Funds of EU in frame of project
26220120021.

\end{document}